\documentclass[conference]{IEEEtran}

\usepackage[pdftex]{graphicx}
\usepackage{amsmath}
\usepackage{amssymb}
\usepackage{color}

\IEEEoverridecommandlockouts

\def\RM{Reed-M\"{u}ller }

\begin{document}

\title{Enhanced Recursive Reed-Muller Erasure Decoding}

\author{
\authorblockN{Alexandre Soro$^1$, J\'er\^ome Lacan$^1$ , Vincent Roca$^2$, Valentin Savin$^3$ and Mathieu Cunche$^{2,4}$}
\authorblockA{$^1$Univ. of Toulouse, ISAE/DISC, 
10 avenue Edouard Belin, BP 54032 - 31055 Toulouse cedex 4 - FRANCE\\
$^2$INRIA Rhone-Alpes, Privatics team, Inovall\'ee, 655 av. de l'Europe, Montbonnot, 38334 St Ismier cedex - FRANCE\\
$^3$CEA-LETI, MINATEC Campus, 17 rue des Martyrs, 38054 Grenoble - FRANCE\\
$^4$INSA-Lyon, Laboratoire CITI 6 Avenue des Arts, 69621 Villeurbanne - FRANCE\\
Email: jerome.lacan@isae.fr, \{vincent.roca,mathieu.cunche\}@inria.fr,valentin.savin@cea.fr}
}

\maketitle

\begin{abstract}
Recent work have shown that \RM (RM) codes achieve the erasure channel capacity. However, this performance is obtained with maximum-likelihood decoding which can be costly for practical applications. In this paper, we propose an encoding/decoding scheme for \RM  codes on the packet erasure channel based on Plotkin construction. We present several improvements over the generic decoding. They allow, for a light cost, to compete with maximum-likelihood  decoding performance, especially on high-rate codes, while significantly outperforming it in terms of speed. 
\end{abstract}

\section{Introduction}

Introduced in 1951 by Mitani \cite{mitani:51} and popularized by Reed \cite{reed:54} and M\"uller \cite{muller:54}, \RM codes are still used in many applications (\textit{e.g.} \cite{dvb_s2}). Their structural properties allow the use of very simple but efficient coding and decoding algorithms. Traditionally, \RM codes were mainly considered for the Gaussian channel because they are able to reach their Maximum-Likelihood (ML) decoding performance with fast soft decision decoding algorithms \cite{dumer,dumer02,JLacanAIAA2007}. But recently, very interesting results have shown that \RM codes achieve the capacity of the erasure channel \cite{CapacityGrouped16,Abbe15}. These results were followed by new contributions on decoding algorithms of these codes \cite{Abbe15,LinearError15}.  

Capacity-achieving erasure codes have interesting practical applications. For example, they are used on higher layers of communication protocols stacks for multicast or real-time transmissions \cite{hanle}. In this context, the data are transmitted into packets and the erasure code is used to generate repair packets from information packets. On the channel, called packet erasure channel, each packet is correctly received or erased.  Classically, packet erasure codes are Fountain codes \cite{rfc5053}, LDPC codes \cite{rfc5170} or MDS (Reed-Solomon-based) codes \cite{rfc5510}. 

The main issue concerning the use of \RM codes on the packet erasure channel is that the traditional decoding algorithm achieving the capacity is based on an inversion (through a Gaussian elimination) of the submatrix of generator matrix corresponding to the received symbols and a matrix-vector multiplication. The gaussian elimination, which has a cubic complexity, is done only once for a codeword of packets but the matrix-vector multiplication is done $z$ times, where $z$ is the size of the packets. This operation can be very costly when the code length increases. 
 
The main contribution of this paper is to propose a sub-quadratic decoding algorithms of \RM codes for the packet erasure channel which obtain performances very close to the capacity. For that, we extend some results presented in \cite{TheseASoro}. The decoding algorithm make use of various structural properties of \RM codes like the Plotkin recursive decomposition \cite{plotkin} and the doubly transitive permutation group. A similar technique was shown to obtain excellent decoding performance on the Gaussian channel \cite{dumer}. Our main contribution is to show that, when used in a different way and combined with additional tools (partial information passing and blank decoding), this techniques yields decoding performance close to maximum likelihood on the packet erasure channel with a very low complexity cost. 


In the next Section, we present the Plotkin construction of Reed-Muller codes and then our algorithm. In Section \ref{results}, we compare the recovering performance and the speed of our decoding algorithm to  the ML decoding with Gaussian Elimination (GE). Finally, we draw some conclusions and perspectives for this algorithm.
\section{Proposed decoding algorithm}
\label{presentation}

\subsection{Introduction to Reed-Muller codes}

Reed-Muller codes defined by parameters $r$ and $m$, where $r\leq m$, are denoted by $RM(r,m)$. The dimension $k$, and the length $n$ of $RM(r,m)$ are given by $k=\sum_{i=0}^{r}{\binom{m}{i}}$ and $n=2^m$.

One way to construct Reed-Muller codes is to use the recursive decomposition introduced by Plotkin. For any $m$, $RM(0,m)$ is defined as the repetition code and $RM(m,m)$ is the identity code. In the general case, for $0<r<m$, any codeword $a\in RM(r,m)$ can be uniquily built from two codewords $u\in RM(r,m-1)$ and $v\in RM(r-1,m-1)$. Precisely, we have:
\begin{multline*}
RM(r,m) = \{(u | u + v), u\in RM(r,m-1),\\
v\in RM(r-1,m-1)\}
\end{multline*}

By using the above decomposition in a recursive manner, it follow that any RM code can be built from repetition and identity codes. This also implies that any $RM(m-1,m)$ code is a Single Parity Check (SPC) code.

Over the erasure channel, a codeword $a$ is received with some erased positions. One can try to decode $u$ and $v$ with $a=(u | u + v)$ recursively. 

For example, if $a\in RM(1,3)$, we have $u\in RM(1,2)$, which is an SPC code, and $v\in RM(0,2)$, which is a repetition code. Therefore, $u$ can be decoded if it has at most one lost position, and $v$ can be decoded if at least one of its position is known. Moreover, decoding new positions on either $u$ or $v$ can reveal some new positions on the other vector, as we have $a_{\frac{n}{2}+i} = u_i + v_i$, for any $0\leq i\leq \frac{n}{2}$. This decoding decomposition will be referred to as the \textit{classical recursive algorithm}. By nature, this algorithm has a  $O(n \log n)$ complexity.
\subsection{Permutation Decoding}
One interesting property of $RM(r,m)$ codes is that they admit the general affine group $GA(m)$ as a permutation group \cite{MWSl77}. In other words, if $a=(a_0,a_1,...,a_{n-1}) \in RM(r,m)$, then $a'=(a_{\Pi(0)},a_{\Pi(1)},...,a_{\Pi(n-1)}) \in RM(r,m)$, where $\Pi(x) = Ax+b$, $A$ is a $m\times m$ binary invertible matrix and $b$ a vector of $(\mathbb{F}_2)^m$. This group is known to be doubly-transitive, \textit{i.e.} there exists a permutation that sends any pair of positions onto any other pair of positions.

It can be observed that the success of the recursive decoding depends on the erasure pattern. Applying some specific permutation on a received word can make successful a decoding that has initially failed. The effects of permutations on the iterative erasure decoding was studied in \cite{hehn} for cyclic and extended cyclic codes, but not for general \RM codes.

\textit{Example :}	
Let us consider that a received pattern of $a\in RM(1,3)$ is $(a_0,\times,\times,\times,\times,a_5,a_6,a_7)$. Only one in $u=(a_0,\times,\times,\times)$ is known thus it can not be decoded. Moreover, no positions of $v$, obtained by summing $u$ and $(u+v)$, are known. Decoding is then impossible.  Using the permutation matrix 
$$A = \left( \begin{matrix} 1&0&0\\ 1&1&0 \\ 0&0&1 \end{matrix} \right) \textrm{ and }b= \left( \begin{matrix} 0\\ 0 \\ 0 \end{matrix} \right),$$ 
the vector to be decoded is then $a'=(a_0,\times, \times, \times, a_6, a_7, \times, a_5)$ and thus $u'=(a_0,\times,\times,\times)$ and $u'+v'=(a_6, a_7, \times, a_5)$. By summing $u'$ and $u'+v'$, we obtain $v'=(a_0+a_6, \times, \times, \times)$.  Since $v'$ is a codeword of the $RM(0,2)$ which is the repetition code of length $4$, $v'_{decoded}=(a_0+a_6, a_0+a_6,a_0+a_6,a_0+a_6)$. By summing $v'_{decoded}$ and the received $u'+v'$, we obtain an updated version of $u'$, $u'_{updated}=(a_0, a_7+a_0+a_6,\times ,a_5+a_0+a_6)$. Since $u'$ belongs to the  $RM(1,2)$ which is a SPC code, we can decode it as $u'_{decoded}=(a_0, a_7+a_0+a_6, a_0+a_5+a_7, a_5+a_0+a_6)$. We can then rebuild the decoded version of $u'+v'$ and finally apply the inverse permutation on $a'$ to obtain the decoded word $a$.	

The key challenge in the previous example is to find a specific permutation, suited for the given erasure pattern. Simulation results have shown that decoding is more efficient by starting the decoding process with the decoding of $v$. As a consequence, the main point is to determine a permutation which maximizes the number of received positions in $v$ for a small complexity cost. 

First we can notice that the vector $b$ is useless. Indeed, for a received word $a=(u | u + v)$, the known positions of $v$ are the positions where $a_i$ and $a_{i + \frac{n}{2}}$ are known. We can observe that the number of known positions of $v$ will not change when the vector $b$  is added to all positions. Without loss of generality, in the following, we restrict ourselves to the set of invertible matrices $A$, \textit{i.e.} the general linear group $GL(m,2)$. For all binary invertible matrices of size $m$, we have to process the images of the $n$ positions. Following \cite{MWSl77} which gives the number of invertible matrices of size $n$, the complexity of testing all possible matrices corresponding to a permutation of size $n$ is roughly $O(n\times 0.29\times 2^{\log^2n})$.

In order to avoid such a prohibitive cost, we propose a lower complexity algorithm. Instead of testing the whole set of invertible matrices, we only select the matrices that leave the $\frac{n}{2}$ leftmost positions, \textit{i.e.} the vector $u$, invariant. It means that their first $m-1$ columns correspond to the first $m-1$ columns of the identity matrix.  
Hence, the number of permutations to be tested is equal to $\frac{n}{2}$, since corresponding matrices are completely determined by their last column, which must be chosen outside the vector subspace spanned by the first $m-1$ columns. For each of these permutations, as the leftmost positions are invariant, only the images of the $\frac{n}{2}$ rightmost positions are processed. The complexity of testing this set of matrices for one permutation of size $n$ is thus $O(\frac{n^2}{4})$. The impact of this permutation selection on the global complexity of the decoding process is discussed in Section \ref{sec:blankDecoding}.


\subsection{Partial Decoding}
Orthogonally to the previous point, we propose a second improvement to the algorithm. One of the weaknesses of the raw recursive decoding algorithm is that, if one recursion fails, the whole decoding fails. This is harmful for the performance as for instance a recursion in $v$ can fail but some new symbols from other recursions may have been recovered. We propose to pass this partial recovered information to the higher levels of the recursion. 

For example, let $a$ be decomposed by $u$ and $v$, which itself is decomposed by $v_1$ and $v_2$. Let the decoding of $v_2$ succeed and $v_1$ fail. $v_2$ may bring some new symbols to $v$, which can themselves bring new symbols to $a$ and then improve the decoding of $u$. Finally, if $u$ reveals new symbols, they can be useful for $v$ and $v_1$ and so forth. Simulation results have shown that most of the time (over 95\%), the number of recursive calls is two, especially for higher levels of recursion, which also means that the use of this mechanism is limited.
\subsection{Blank Decoding}
\label{sec:blankDecoding}
Unlike the previous improvements enhancing the decoding capability, "blank decoding" is focused on speed. 

To explain this mechanism, let us consider a packet erasure code generating $n-k$ repair packets from $k$ information packets. By assuming that all packets have size $z$ bits, the encoding of the packet code can be seen as the parallel encoding of $z$ binary codewords of size $n$. Each one of the $n$ transmitted packets contains exactly one bit of each of the $z$ encoded vectors. The interest of this scheme is that the erased packets produce the same erasure pattern on all the binary codewords. In the decoding process of the packet code, this property is used to pool together operations that are common to the $z$ decodings of the binary codewords.   


Actually, before each packet block decoding, a blank block (\textit{i. e.} a block with packets of null size) decoding is processed with the received loss pattern. During this process, the decoder searches the best permutation to apply and the best partial information passing strategy, according to the algorithms described above and records it. If this decoding succeeds, the actual decoding of the $z$ binary codewords are simply done by following the path traced by the blank pattern. If the blank decoding fails, a partial decoding may however be possible. A Gaussian Elimination may also be attempted, also in a blank mode, either after or instead of this partial decoding.

As seen previously, the main point is to find the permutations recursively. Consequently the complexity of this blank decoding is roughly $O(n^2\log n)$, processed only once per packet block. This cost must be compared to the decodings of the $z$ binary codewords (each one having a complexity $O(n\log n)$). The impact of blank decoding on the global decoding complexity of a packet block is evaluated in next section. 


\section{Results}
\label{results}
\begin{figure}[htb]
  \begin{center}
   \includegraphics[width=.49\textwidth]{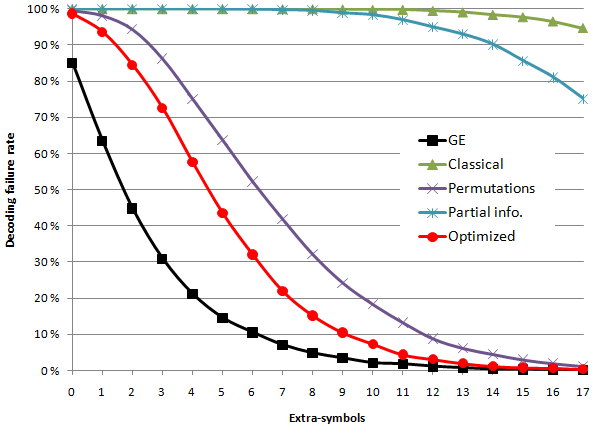}
    \end{center}
    \caption{Decoding failure rate of GE and recursive algorithms in terms of extra-symbols for $RM(3,7)$}
   \label{fig:ineff}
\end{figure}

We have compared our proposal with maximum likelihood decoding using Gaussian Elimination (GE) on an Intel Core 2 Extreme @3.06Ghz on Mac OS X 10.6 in 64-bit mode. Fig. \ref{fig:ineff} shows the decoding failure probabilities for $RM(3,7)$ code ($k=64$, $n=128$), in terms of extra-packets. By extra-packets, we mean the number of received symbols above the source block size $k$, which corresponds to  the erasure channel capacity. We compare ML, which is optimal, with our algorithm and the classical recursive algorithm. In addition, we provide results for the recursive algorithm with only permutation selection and only partial information passing.
We can see that our algorithm performs well compared to the optimal GE and requires only 3 extra-symbols in average to match the performances of GE. On the opposite, the raw recursive algorithm is not able to recover anything with up to 20\% of extra-symbols. For $RM(3,7)$, the decoding speed of both algorithms is provided in Table \ref{tab:rm}. For the sake of completeness,  the encoding speed for the RM(3,7) code is in the order of 4Gbps, and the decoding speed of a Reed-Solomon code with the same parameters is around 350Mbps \cite{rfc5510}.

In Table \ref{tab:rm}, we provide a comparison of the decoding speed and the average overhead between GE and our algorithm for various \RM codes. All results are provided for packets of 1500 bytes.
\begin{table}[ht]
\begin{center}
\begin{tabular}{|r|c|c|c|c|c|}
\hline
		 &  Recursive & Gaussian Elim. & Recursive & ML \\
		& speed	& speed & overhead & overhead \\
\hline
RM(3,6)	& 2021 Mbps &  842 Mbps & 5.41\% & 5.06\% \\
\hline
RM(3,7)	& 1073 Mbps &  544 Mbps & 8.59\% & 4.75\% \\
\hline
RM(4,7)	& 2393 Mbps &  381 Mbps & 3.45\% & 2.79\% \\
\hline
RM(4,8)	& 1363 Mbps &  215 Mbps & 9.08\% & 1.44\% \\
\hline
RM(5,8)	& 2774 Mbps &  181 Mbps & 2.44\% & 1.17\% \\
\hline
RM(5,9)	& 1783 Mbps &  85 Mbps & 9.23\% & 0.45\% \\
\hline
RM(6,9)	& 3291 Mbps &  80 Mbps & 1.90\% & 0.47\% \\
\hline
RM(6,10)	& 3486 Mbps &  44 Mbps & 8.05\% & 0.18\% \\
\hline
\end{tabular}
\end{center}
\caption{Decoding speed and overhead for various RM codes}
\label{tab:rm}
\end{table}

For the $RM(6,9)$ ($k=466$, $n=512$) code, we compare the decoding speed of GE with our implementation. We provide results for packets of $50$, $500$ and $1500$ bytes which represent classical VoIP, median Internet and LAN data units, in Table \ref{tab:ineff}.

\begin{table}[ht]
\begin{center}
\begin{tabular}{|r|c|c|c|c|}
\hline
Packet Size		 &  Recursive Alg. & Gaussian Elim. & Speed Ratio\\
\hline
50 bytes	& 546 Mbps &  8 Mbps & 66x \\
\hline
500 bytes	& 2463 Mbps &  51 Mbps & 49x \\
\hline
1500 bytes	& 3291 Mbps &  80 Mbps & 41x \\
\hline
\end{tabular}
\end{center}
\caption{Decoding speed for $RM(6,9)$ code with $5\%$ of extra-symbols}
\label{tab:ineff}
\end{table}
For this code, the speed up ratio between our code and the GE is between 40x and 60x, depending on the packet size. This variation can be explained by the fact that the smaller is the packet size, the longer is the pre-decoding phase compared to the actual decoding. This pre-decoding is more complex for GE (a cubic complexity matrix inversion) than the blank decoding for the recursive algorithm. Note, however, that the blank decoding represents $86\%$, $38\%$ and $17\%$ of decoding time for recursive decoding for packet sizes of $50$, $500$ and $1500$ bytes. It is worth pointing out that for the recursive decoding, only the encoded vector is recovered. We do not take into account the final step which consists in recovering the source symbols from the whole encoded vector, which is in the magnitude of $8$ Gbps for this code, and thus far faster than the decoding.\\

In high code rates scenarios, our algorithm is able to decode with only $1\%-2\%$ more symbols than the optimal GE decoding. Depending on the target application, one may be interested to use this decoding algorithm, as it allows a ratio of 5x - 20x decoding speed compared to GE.
\section{Conclusions}
\label{conclusions}
In this contribution, we have presented a recursive decoding algorithm for Reed-Muller codes based on Plotkin construction. We have shown that the permutation selection and the blank decoding dramatically improve the performances of the decoder. In many scenarios, our algorithm can achieve decoding performance close to the capacity, while achieving speeds of several Gbps. Because of their high decoding speed, \RM codes are an interesting alternative for the packet erasure channel, when the decoding cost is a bottleneck (\textit{e.g.} in low power devices and real-time applications). 
\bibliographystyle{IEEEtran}
\bibliography{rm, RM_aiaa}

\begin{thebibliography}{10}
\providecommand{\url}[1]{#1}
\csname url@samestyle\endcsname
\providecommand{\newblock}{\relax}
\providecommand{\bibinfo}[2]{#2}
\providecommand{\BIBentrySTDinterwordspacing}{\spaceskip=0pt\relax}
\providecommand{\BIBentryALTinterwordstretchfactor}{4}
\providecommand{\BIBentryALTinterwordspacing}{\spaceskip=\fontdimen2\font plus
\BIBentryALTinterwordstretchfactor\fontdimen3\font minus
  \fontdimen4\font\relax}
\providecommand{\BIBforeignlanguage}[2]{{%
\expandafter\ifx\csname l@#1\endcsname\relax
\typeout{** WARNING: IEEEtran.bst: No hyphenation pattern has been}%
\typeout{** loaded for the language `#1'. Using the pattern for}%
\typeout{** the default language instead.}%
\else
\language=\csname l@#1\endcsname
\fi
#2}}
\providecommand{\BIBdecl}{\relax}
\BIBdecl

\bibitem{mitani:51}
N.~Mitani, ``{On the Transmission of Numbers in a Sequential Computer},''
  \emph{National Convention of Inst. Elect Engineers of Japan}, 1951.

\bibitem{reed:54}
I.~Reed, ``{Decoding of Reed-Muller Codes with a Large Number of Errors},''
  \emph{IEEE Transactions on Information Theory}, no.~4, 1954.

\bibitem{muller:54}
D.~Muller, ``{Application of Boolean algebra to switching circuit design and to
  error detection},'' \emph{IRE Trans. Electron. Comput.}, no.~3, 1954.

\bibitem{dvb_s2}
ETSI, ``{{ Digital Video Broadcasting (DVB) ; Second generation framing
  structure, channel coding and modulation systems for Broadcasting,
  Interactive Services, News Gathering and other broadband satellite
  applications (DVB-S2), EN 302 307}},'' 2004.

\bibitem{dumer}
I.~Dumer and K.~Shabunov, ``{Soft-decision decoding of Reed-Muller codes:
  recursive lists.}'' \emph{IEEE Transactions on Information Theory}, vol.~52,
  pp. 1260--1266, 2006.

\bibitem{dumer02}
------, ``{Recursive and permutation decoding for Reed-Muller codes.}''
  \emph{IEEE International Symposium on Information Theory}, p. 146, 2002.

\bibitem{JLacanAIAA2007}
Y.~Oster, J.~Lacan, and A.~Duverdier, ``{Benchmark of {R}eed-{M}üller {C}odes
  for Short Packet Transmission},'' in \emph{Proceedings of the 25th AIAA
  International Communications Satellite Systems Conference}, April 2007.

\bibitem{CapacityGrouped16}
S.~Kumar, S.~Kudekar, M.~Mondelli, H.~D. Pfister, E.~Sasoglu, and R.~L.
  Urbanke, ``{Reed-Muller Codes Achieve Capacity on Erasure Channels},'' vol.
  abs/1601.04689, january 2016.

\bibitem{Abbe15}
E.~Abbe, A.~Shpilka, and A.~Wigderson, ``{Reed-Muller Codes for Random Erasures
  and Errors},'' in \emph{Proceedings of the Forty-Seventh Annual ACM on
  Symposium on Theory of Computing}, ser. STOC '15, 2015.

\bibitem{LinearError15}
R.~Saptharishi, A.~Shpilka, and B.~L. Volk, ``{Decoding high rate Reed-Muller
  codes from random errors in near linear time},'' vol. abs/1503.09092, 2015.

\bibitem{hanle}
C.~H\"anle, ``Feasibility study of erasure correction for multicast file
  distribution using the network simulator ns-2,'' \emph{IEEE Military
  Communications Conference}, vol.~3, pp. 1260--1266, 1998.

\bibitem{rfc5053}
M.~Luby, A.~Shokrollahi, M.~Watson, and T.~Stockhammer, ``{Raptor Forward Error
  Correction Scheme for Object Delivery},'' RFC 5053 (Proposed Standard), IETF,
  Oct. 2007.

\bibitem{rfc5170}
V.~Roca, C.~Neumann, and D.~Furodet, ``{Low Density Parity Check (LDPC)
  Staircase and Triangle Forward Error Correction (FEC) Schemes},'' RFC 5170
  (Proposed Standard), IETF, Jun. 2008.

\bibitem{rfc5510}
J.~Lacan, V.~Roca, J.~Peltotalo, and S.~Peltotalo, ``{Reed-Solomon Forward
  Error Correction (FEC) Schemes},'' RFC 5510 (Proposed Standard), IETF, Apr.
  2009.

\bibitem{TheseASoro}
A.~Soro, ``M\'ecanismes de fiabilisation pro-actifs,'' Ph.D. dissertation,
  University of Toulouse, december 2010.

\bibitem{plotkin}
M.~Plotkin, ``Binary codes with specified minimum distance,'' \emph{IRE
  Transactions on Information Theory}, vol.~6, pp. 445--450, 1960.

\bibitem{MWSl77}
F.~J. MacWilliams and N.~J.~A. Sloane, \emph{{The Theory of Error-Correcting
  Codes}}.\hskip 1em plus 0.5em minus 0.4em\relax North-Holland, 1977.

\bibitem{hehn}
T.~Hehn, O.~Milenkovic, S.~Laendner, and J.~B. Huber, ``{Permutation Decoding
  and the Stopping Redundancy Hierarchy of Cyclic and Extended Cyclic Codes.}''
  \emph{IEEE Transactions on Information Theory}, no.~12, pp. 5308--5331.

\end{thebibliography}

\end{document}